# Silicon-CMOS Compatible In-Situ CCVD Grown Graphene Transistors with Ultra-High On/Off-Current Ratio

*Pia Juliane Wessely, Frank Wessely, Emrah Birinci,*

*Karsten Beckmann, Bernadette Riedinger[a], Udo Schwalke*

*Technische Universität Darmstadt, Schlossgartenstrasse 8, 64289 Darmstadt, Germany*

[a]*Fraunhofer-Institut für Werkstoffmechanik, Wöhlerstrasse 11, 79108 Freiburg, Germany*

*Phone: +49 6151 163931; Fax: +49 6151 165233; E-mail: pj.wessely@iht.tu-darmstadt.de*

ABSTRACT - By means of catalytic chemical vapor deposition (CCVD) in-situ grown monolayer graphene field-effect transistors (MoLGFETs) and bilayer graphene transistors (BiLGFETs) are realized directly on oxidized silicon substrate without the need to transfer graphene layers. In-situ grown MoLGFETs exhibit the expected Dirac point together with the typical low on/off-current ratios.  In contrast, BiLGFETs possess unipolar p-type device characteristics with an extremely high on/off-current ratio up to $1\times10^7$. The complete fabrication process is silicon CMOS compatible. This will allow a simple and low-cost integration of graphene devices for nanoelectronic applications in a hybrid silicon CMOS environment.

# 1. INTRODUCTION

A monolayer of graphene consists of carbon atoms which are arranged in a quasi planar honeycomb lattice structure. It is a true 2D material which has been first synthesized by A. Geim and K. Novoselov [1] in 2004. Meanwhile researchers distinguish between monolayer graphene, bilayer graphene, trilayer graphene und fewlayer graphene (i.e. five to ten stacked graphene sheets). Currently there are several different manufacturing methods to produce graphene. Most of them have in common that a subsequent transfer of the graphene layer onto a suitable substrate is required after synthesis [2].

Exfoliation is the method invented by A. Geim and K. Novoselov used to realize graphene layers for the first time [1]. By this means single to fewlayer graphene films can be deposited on a substrate wafer. However, size and position of the graphene flakes varies randomly. In addition to this difficulty, adsorbed molecules like $O_2$ and $H_2O$ often accumulate at the interface between graphene and the substrate surface [3]. Another possibility to produce graphene films is the use of CVD based methods to grow graphene on metallic substrates like copper or nickel [4]. Although large area of graphene films can be produced [5] the disadvantage for integrated electronic applications is the need to transfer and align the produced graphene layer to the silicon substrate, for example [6]. Very recently a modified CVD-based approach has been reported which relies on the scalable synthesis of graphene on patterned Ni-dots [7]. Nevertheless, after the growth of the graphene sheets on the Ni dots the transfer of the graphene layers is still necessary. In order to avoid graphene transfer, epitaxial graphene grown on silicon carbide (SiC) has been proposed by de Heer and Berger [8]. Using this method fairly large graphene sheets can be realized on a SiC wafer without the need to transfer. However, when comparing with conventional silicon processing, this method is more expensive because of the SiC substrate. Furthermore, the process requires extraordinary high growth temperatures of about 1400°C and is therefore not compatible with conventional silicon CMOS processing.

In order to avoid all the above mentioned drawbacks we have developed an in-situ CCVD growth method of graphene on oxidized silicon wafers. In-situ means, that graphene films are grown directly on the wafer at its final position, so that subsequent transfer and alignments are obsolete. Using a metallic catalyst seed, we are able to grow mono-, bi- and fewlayer graphene films directly on silicon-dioxide ($SiO_2$) covered Si wafers at moderate growth temperatures of 800 - 900°C by means of CCVD from a methane feedstock [9]. During growth the graphene layer extends a few microns from the catalyst onto the oxidized wafer surface. When using a suitable device layout, the in-situ grown graphene film can be used as back-gated field effect device material contacted directly via the catalytic source/drain areas (cf. Fig. 1) for electrical characterization. Furthermore, by adjusting the CCVD process conditions we can determine the number of grown graphene layers giving us the unique capability to investigate the physical and electrical properties of various in-situ grown graphene films at the device level. We found that large area bilayer graphene field effect transistors exhibit an extremely high on/off-current ratio of $1 \times 10^7$ at room temperature.

## 2. FABRICATION AND STRUCTURAL ANALYSIS

Starting with an oxidized silicon wafer containing 100nm thick silicon dioxide several lithography steps follow until a structured liftoff resist remains on the wafer surface. In preparation for CCVD nm-thin aluminum and nickel layers are evaporated over the whole substrate surface and are structured via liftoff. During annealing the aluminum partially transforms into aluminum-oxide ($Al_xO_y$) while the nickel (Ni) layer generates several nickel nanoclusters at the perimeter of the catalyst system (see Fig. 1a). Nickel nanoclusters act as seed for the subsequent graphene layer growth in the methane-based CCVD process. Using the CCVD process several hundred graphene devices are realized simultaneously across one 2" wafer and are directly functional after the CCVD growth. Fig 1b shows a scanning electron microscopy picture of the device area. These graphene devices possess a well defined channel length in the range of 1.6µm to 5µm while the channel width varies randomly from approximately 0.1µm to several microns, depending on local growth conditions. The in-situ grown graphene layers extend only a few microns on the $SiO_2$ surface and therefore do not always fill up the maximum designed channel width. The magnification in Fig. 1c shows a single graphene field effect device with a channel length of 1.6µm and a maximum channel width of 5µm. In addition to the grown graphene layers on the oxide surface additional carbon deposits are present on top of the catalytic areas as evident from the enlargement shown in Fig. 1d. Besides graphene layers, also carbon nanotubes (CNTs) and additional carbon deposits have been found on top of the metal catalyst. Detailed examination by means of scanning electron microscopy indicates that the spread of the CNTs and the additional carbon deposits occur exclusively on the top of catalyst source/drain (S/D) areas.

The number of grown stacked graphene layers depends on the adjusted process parameters, e.g. temperature and gas mixture. First, measurements of topography are performed by atomic force microscopy (AFM) as shown in the examples in Figs. 1e and 1f. From the step-height measurement in Fig. 1f a value of 0.6nm is deduced at the graphene to silicon dioxide transition indicating the presence of bilayer graphene.

Subsequently, Raman spectroscopy is performed within the channel region in between the catalytic areas (cf. Fig. 1d) using a Renishaw spectrometer at 633 nm at room temperature. In Raman spectra of graphene three main peaks can be determined. The D and G peak always appear in Raman spectra of $sp^2$-carbon materials. G and D peak at 1338 $cm^{-1}$ and 1578 $cm^{-1}$ respectively, represent the graphitic $sp^2$-structure (G peak) and the defects in the graphene lattice, as holes and edges (D peak). The 2D band around 2650 $cm^{-1}$ is known as the second order of the D peak. The shape of this peak is characteristic for the number of stacked graphene layers [10]. Comparing Fig. 2 with Raman data from A. Ferrari [10] suggests the presence of monolayer and bilayer graphene in our sample. The characteristic Raman G and D as well as the 2D band are located at similar Raman shift positions found by Ferrari and exhibit the expected shape.

However, a difference in the I(2D)/I(G) ratio can be observed. The influences of the substrate on the Raman spectra also need to be considered in the interpretation as already addressed by [11] and [12]. Raman spectra which have been measured on graphene prepared by micromechanical cleavage and subsequent transfer to silicon dioxide [12] deviate significantly in the I(2D)/I(G) ratio from the Raman spectra of in-situ grown CCVD graphene shown in Fig. 2. These differences indicate strong interactions of graphene with underlying silicon dioxide [11, 12] which are due to the in-situ growth at moderate temperatures. In contrast, for externally grown graphene with subsequent transfer such intensive interactions between graphene and silicon dioxide are largely reduced in presence of adsorbed molecules, e.g. $O_2$ and $H_2O$.

To complete these studies, additional transmission electron microscopy (TEM) studies have been performed on CCVD fewlayer graphene samples followed by a fourier analysis in order to determine the crystalline properties of the carbon multilayer more in detail. The observed interplanar spacing of 3.5Å [9] is in fact a strong evidence for the existence of graphene grown by means of CCVD.

3. RESULTS AND DISCUSSION

The electrical characterization of the graphene devices is performed using a Keithley SCS 4200 semiconductor analyzer. The metal catalyst areas are directly used as source and drain contacts (cf. Fig 1a). However, carbon deposits, like carbon nanotubes on the metal catalyst surface (see Fig. 1d), are expected to increase the contact resistance. Fig. 3a shows the transfer characteristic of the same **MonoLayer Graphene Field Effect Transistor** (MoLGFET), on which the previously discussed Raman spectrum of monolayer graphene has been obtained in the channel region. The characteristic Dirac-point at $V_{BG}$= -4V confirms the co-existence of hole and electron conduction together with the typical low on/off-current ratios, as expected for monolayer graphene [13]. However, a slightly unsymmetrical current-voltage characteristic is noted in Fig. 3a, leading to different on/off-current ratios of 16 for hole-conduction and 8 for electron-conduction, respectively. Obviously, hole-conduction is preferred in our in-situ CCVD grown graphene, which appears typical for CVD-grown graphene according to Hall-effect measurements reported from other groups [14]. Furthermore Fig. 3b shows the corresponding output characteristic which exhibits the typical three region behavior of a large area monolayer graphene transistor as described in [13], confirming the existence of in-situ CCVD grown monolayer graphene and substantiate the prediction of the Raman spectra.

Since monolayer graphene is considered to be a zero bandgap material [13], large area monolayer graphene (MoLG) layers are not appropriate as channel material to be used in field-effect devices for digital applications [2]. In order to realize MoLG semiconducting materials one possibility is to constrict the lateral dimensions of the graphene layer to a few nanometers, forming the so called graphene nanoribbons (GNRs). GNRs exhibit a bandgap up to approximately 0.5eV [2], depending on the ribbon width. However, sub 10nm dimension are difficult to realize and additional difficulties related to line edge roughness [15] and edge defects which will degrade the performance of graphene devices.

To avoid problems with GNR-fabrication, one can use bilayer graphene (BiLG) instead of monolayer graphene for device fabrication. Despite the large area a bandgap can be induced in BiLG by applying an electrical field perpendicular to the layer [16, 17]. The transfer characteristics of a typical CCVD in-situ grown **BiL**ayer **G**raphene **F**ield **E**ffect **T**ransitor (BiLGFET) is shown in Fig. 4 together with the output characteristic presented in the inset of Fig. 4. In contrast to MoLGFETs these BiLGFETs exhibit an ultra-high on/off-current ratio of $1 \times 10^7$ which is to our knowledge the highest reported value for in-situ CCVD grown BiLGFETs today. The transfer characteristic is consistent with the output characteristic, displaying a unipolar, p-type device behavior. Such a unipolar device characteristics requires both, a substantial bandgap and a kind of "carrier selection mechanism", since only holes are the predominant carriers in a p-type field-effect device. P. Avouris et al [18] achieved ambipolar, double gated bilayer graphene FETs with improved on/off-current ratios of 100 to 2000. The bandgap is exclusively induced by the applied electrical field. In contrast, for

CCVD BiLGFETs the fairly low gate field strength of 1.5 MV/cm (i.e. 0.15V/nm) perpendicular to the graphene layer is insufficient to create a significant bandgap [16] to explain these device characteristics. Accordingly, we suspect that additional effects, like intensive interactions between bilayer graphene and silicon dioxide are responsible to further enhance the bandgap. Such intensive interactions may develop during the growth of the bilayer graphene on the silicon dioxide at moderate temperatures under well defined ambient conditions within a CVD chamber. In fact, in a similar fashion graphene-substrate interactions have been used to explain the substrate-induced bandgap opening in epitaxial graphene [19]. The selection of the carrier type (i.e. holes in this case) may be facilitated by additional doping and/or Schottky-barrier effects. Since substantial amounts of atomic hydrogen are generated from the decomposition of $CH_4$ during CCVD processing, it is likely that H-atoms adsorb on the graphene surface or may be incorporated within the graphene bilayer. As a result, effects on the electronic properties such as increasing the bandgap are expected [20]. Furthermore, in view of the intensive graphene-substrate interactions, we suspect that atomic hydrogen, which is also known to passivate interface traps in $SiO_2$ [21], plays an important role in the carrier type selection. In case of vacuum-degassed graphene and in the absence of atomic hydrogen, n-type behavior has been reported [22] which is due to Fermi level pinning in the presence of interface states. In contrast, when large amounts of hydrogen are present during in-situ graphene growth $SiO_2$ interface states are passivated and the p-type behavior is preserved in our CCVD grown graphene [22]. However, to which extent the p-type characteristics of the BiLGFETs is primarily related to atomic hydrogen doping, similarly to the tunable polarities of CNTs with oxygen and metal atoms [23], needs further investigation. Finally, Schottky-barrier formation at the metal-semiconductor interface [24] of the nickel-graphene contacts needs also to be considered to explain the unipolar behavior of the BiLGFETs.

4. CONCLUSION

The combination of AFM examination, Raman spectroscopy as well as extensive electrical characterization of graphene structures on $SiO_2$ confirms the suitability of this novel in-situ CCVD growth process. MoLGFET show the expected characteristic Dirac-point and a typical low on/off-current ratio of 16. In contrast, BiLGFETs exhibit ultra-high on/off-current ratios of $1 \times 10^7$ exceeding previously reported values by several orders of magnitude. We explain the improved device characteristics by a combination of effects, in particular graphene-substrate interactions, hydrogen doping and Schottky-barrier effects at the S/D contacts as well. The transfer characteristic is consistent with the output characteristic, displaying a clear unipolar p-type device behavior. With this novel fabrication method hundreds of large scale BiLGFETs are realized simultaneously on one 2'' wafer by in-situ CCVD grown BiLG in a silicon CMOS compatible process.

ACKNOWLEDGMENTS This research is part of the ELOGRAPH project within the ESF EuroGRAPHENE EUROCORES programme.

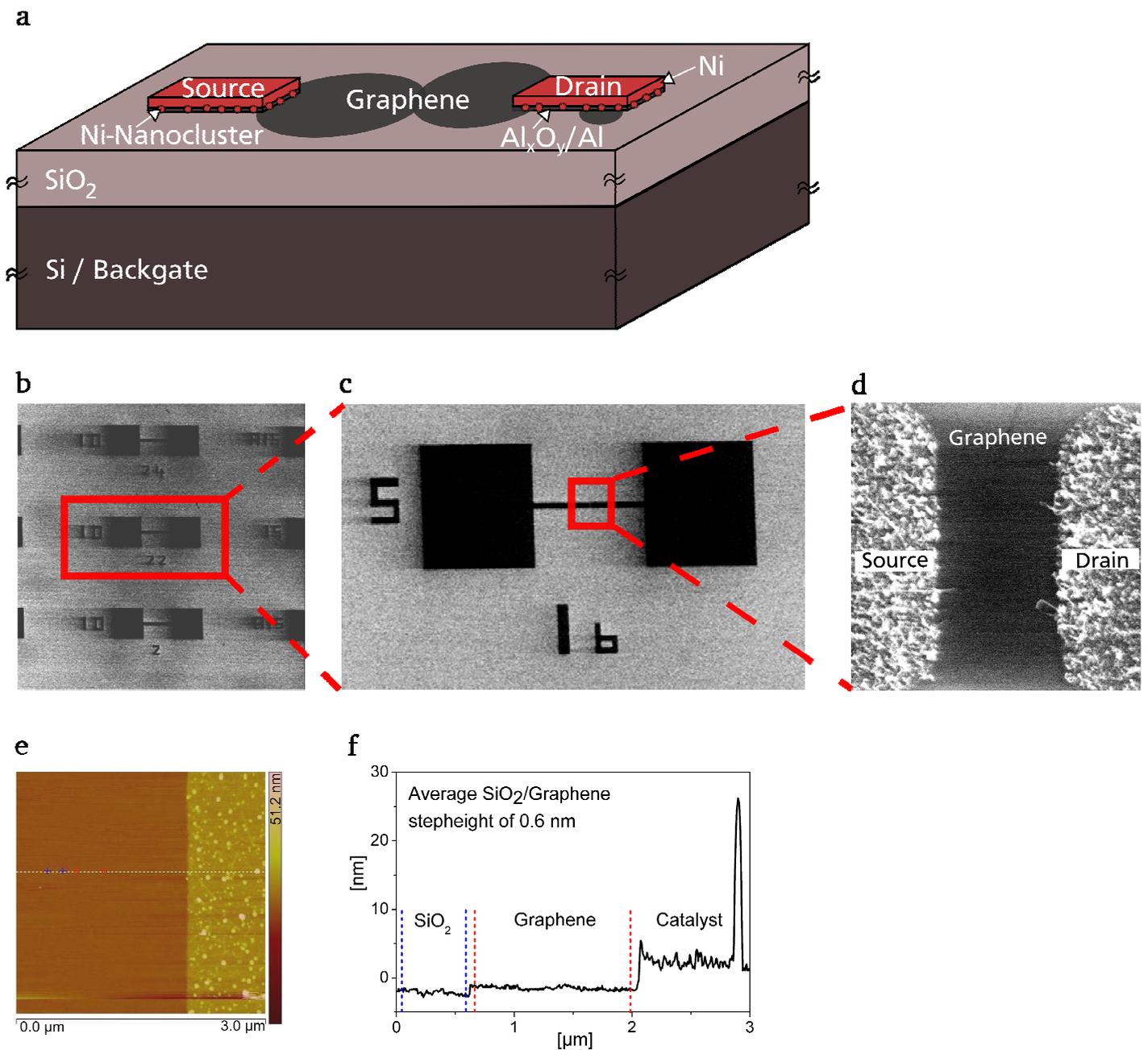

Figure 1: **a,** Schematic representation of a graphene device produced by means of in-situ CCVD using a nickel catalyst. **b,** Scanning electron microscopy examination of a contact pad area. The contact pad size is 100µm x 100µm. Hundreds of graphene field-effect devices are realized simultaneously on one 2'' wafer. **c,** Magnification of a graphene field effect device with a channel length of 1.6µm and a maximum technological channel width of 5µm in this case. **d,** Further magnification of a graphene field effect device at the channel

region. Source and drain regions are connected by a graphene layer. Additionally carbon deposits at source and drain region can be seen. **e,** AFM Measurement of the graphene layer on a silicon dioxide surface growing form the catalyst area on the right side of the picture. **f**, Corresponding stepheight analysis by average height value calculated at the red and blue lines. Stepheight of 0.6 nm corresponds to bilayer graphene.

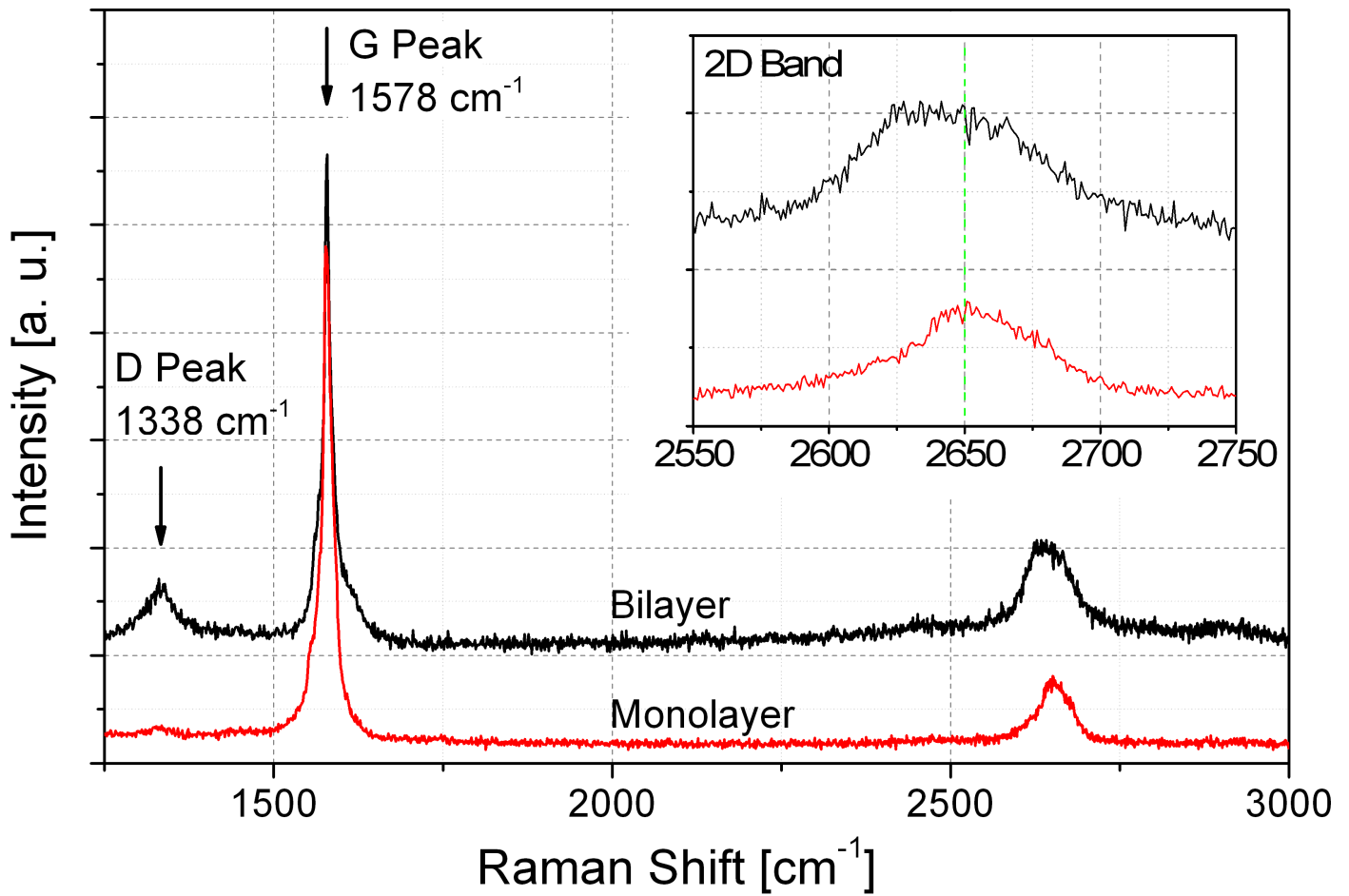

Figure 2: Raman spectra of in-situ CCVD grown monolayer and bilayer graphene. The Raman-measurements are performed using a Renishaw spectrometer at 633nm at room temperature. The inset shows the magnification of the 2D band in graphene. Comparing these results with [9], [10] and [11] indicates the presence of monolayer and bilayer graphene.

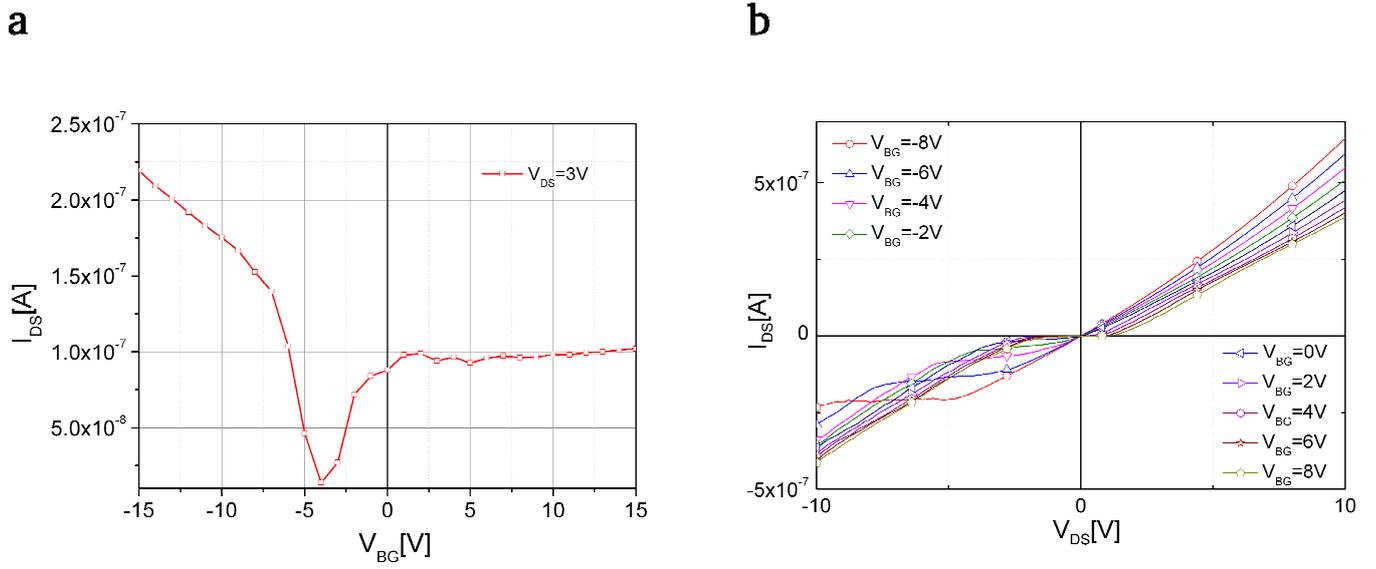

Figure 3: **a,** Current vs. gate-voltage characteristic of monolayer graphene. The Dirac point indicates simultaneous electron and hole conduction. **b,** The output characteristic of a large area monolayer graphene transistor (MoLGFET) exhibits typical three region behavior [12].

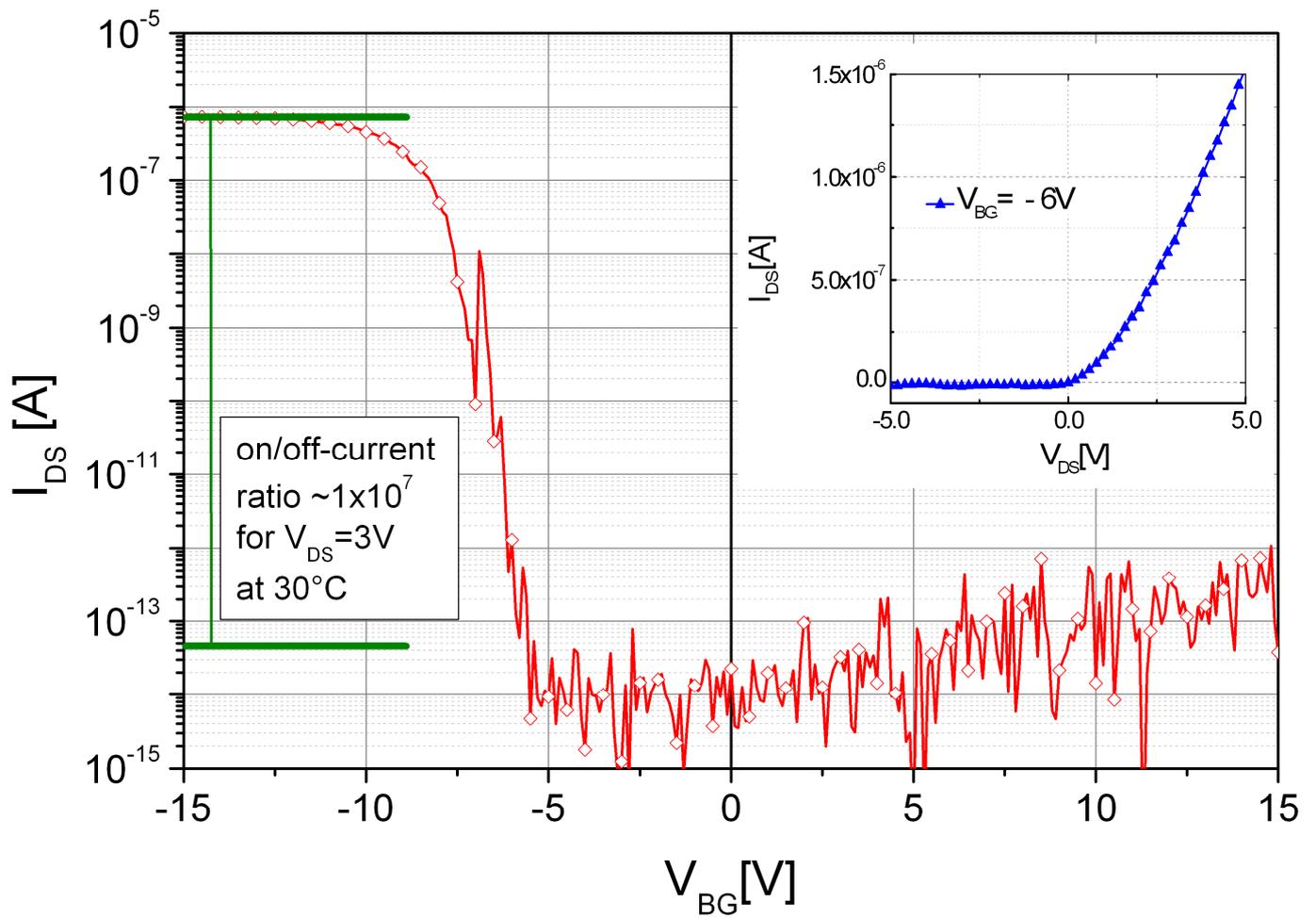

Figure 4: Current vs. gate-voltage characteristic of a bilayer graphene transistor. The backgate voltage $V_{BG}$ is swept from -15V to 15V while a constant voltage $V_{DS}$ of 3V is applied between drain and source. Extremely high on/off-current ratios of $1\times10^7$ at 30°C are observed for the bilayer graphene transistor. *Inset:* Output characteristic of a bilayer graphene field effect transistor at backgate voltages $V_{BG}$ = -6V while the drain/source voltage $V_{DS}$ is swept from -5V to 5V.